\documentclass[11pt]{article}
\makeatletter
\@addtoreset{equation}{section}

\makeatother

\begin{document}

\begin{titlepage}

\renewcommand{\thefootnote}{\fnsymbol{footnote}}

\hfill\parbox{4cm}{hep-th/0507091}
\vspace{15mm}
\baselineskip 9mm
\begin{center}
{\LARGE \bf D3 instantons in  Calabi-Yau orientifolds with(out) fluxes}

\end{center}

\baselineskip 6mm
\vspace{10mm}
\begin{center}
Jaemo Park\footnote{\tt jaemo@physics.postech.ac.kr} 

\vspace{3mm}
{\sl  Department of Physics, Pohang University of Science and Technology
\newline
 Pohang 790-784,
Korea}

\end{center}

\thispagestyle{empty}

\vfill
\begin{center}
{\bf Abstract}
\end{center}
\noindent
We investigate the instanton effects due to  D3 branes wrapping a four-cycle 
in a Calabi-Yau orientifold with D7 branes. We study the condition for the 
nonzero superpotentials from the D3 instantons. For that matter we work out the 
zero mode structures of D3 branes wrapping a four-cycle both in the presence of 
the fluxes and in the absence of the fluxes. In the presence of the fluxes, the 
condition for the nonzero superpotential could be different from that without 
the fluxes. We explicitly work out a simple example of the orientifold of 
$K3 \times T^2/Z_2$ with a suitable flux to show such behavior. The effects of D3-D7
sectors are interesting and give further constraints for the nonzero superpotential.
In a special configuration where D3 branes and D7 branes wrap the same four-cycle, 
multi-instanton calculus of D3 branes 
could be reduced to that of a suitable field theory. 
The structure of D5 instantons in Type I theory is briefly discussed. 
\vspace{20mm}
\end{titlepage}

\vspace{1cm}

\section{Introduction}

Understanding the nonperturbative corrections to the superpotentials is 
an interesting topic in string theory. With the recent progress in understanding 
the flux compactification, it might be interesting to pursue this issue in this 
context. Especially, KKLT\cite{KKLT} type scenario crucially needs such nonperturbative 
corrections. So it would be interesting to figure out the conditions where 
such corrections exist. One convenient starting point for the flux compactification 
is to consider the Type IIB orientifold on Calabi-Yau manifolds with three-form 
fluxes and D7-branes\cite{GKP}. The nonperturbative corrections in this set up are due to 
D3-branes. In order to understand such instanton effects, we need the physical gauge approach 
as developed in \cite{Moore, Witten, Saulina, Park1, Park2}. 
Furthermore the instanton effects due to D3 branes are not well 
understood in the Calabi-Yau compactification even without the fluxes. 

Here we initiate the investigation of the D3 instanton effect in the Calabi-Yau 
orientifolds. In this paper we are mainly interested in the basic structure 
of D3 instantons and the conditions for the nonzero superpotentials. Much of the 
discussion is devoted to the fermion zero modes on  D3 branes wrapping a four cycle
in a Calabi-Yau orientifold, relegating the further discussion of the specific 
examples to future work. We find  the conditions for the nonzero superpotentials 
in the presence of the fluxes and in the absence of the fluxes. The nonzero conditions 
in the presence of the fluxes are different from those in the absence of the fluxes.
Related examples were studied in \cite{Gorlich, Mayr} and the related M5 brane instanton 
effects were 
discussed in \cite{Saulina2, Trivedi, Kallosh, Aspinwall}. 
We consider one simple orientifold of $K3 \times T^2/Z_2$ with D7 branes
as an example where the modification of the nonzero conditions for the superpotentials 
occur due to the presence of the fluxes. We show that this orientifold can have 
nonzero superpotentials due to D3 instantons in the presence of the fluxes while 
without the fluxes there would be no superpotentials arising from D3 instantons. 
Along with this development we find many interesting facts as well. Especially 
the D3-D7 sectors are important in understanding the structure of the D3 instantons. 
When D3 branes and D7 brane wrapping the same four cycle, D3 instantons could be reduced 
to the usual field theory instantons. Then we can borrow the results of the multi-instanton 
calculus of the field theory to study the multi-instanton effects of D3 branes, which 
are not well understood so far. In other case of D3-D7 sectors, we find D1 string sector
at the intersection of D3 brane and D7 brane. The effect of the D1 string is similar to 
that of the heterotic string instanton effect as considered in \cite{Witten, Park1}
and this gives rise 
to further restrictions on the nonzero conditions for the superpotentials. 

The content of the paper is as follows. In section 2, we consider the fermion 
zero modes on the D3 brane world volume to figure out the nonzero conditions for the 
superpotentials due to the D3 instantons wrapping a four-cycle in the 
Calabi-Yau orientifolds with D7 branes. In section 3, we consider the same 
problem in the presence of the fluxes. The nonzero conditions are dependent 
on the fluxes and we are mainly interested in the simple orientifold 
of $K3 \times T^2/Z_2$ with D7 branes. We show that there indeed nonzero superpotentials 
generated due to D3 instantons. In section 5, by the similar method we work out 
the condition where the contribution of the D5 brane instanton effects to the 
superpotential in Type I theory is nonzero. After this work has finished, we are aware of the 
work\cite{Kallosh2} which deals with the similar problem.

\vspace{1cm}

\section{Zero mode analysis in the absence of the fluxes}

\vspace{5mm}

We are mainly interested in the Calabi-Yau orientifold with D7 branes
and (instantonic) D3 branes.
We will consider two possibilities for D3-D7
systems. We assume that the normal directions to D7
branes are $x^8, x^9$ directions and Calabi-Yau manifold spans $x^4$ to $x^9$. 
In one case D3 brane world volume
directions span $x^4, x^5, x^6, x^7$ directions so that within the
Calabi-Yau manifold, the transverse directions of D3 and D7 branes 
coincide. This is T-dual to D5-D9 configurations if $x^8, x^9$ directions are 
compactified on $T^2$, which are small 
instanton ones. (In a nontrivial geometry of Calabi-Yau, the coordinates 
$x^4 \cdots x^9$ are local coordinates near the D3 brane.) Another
possibility
is that the D3 brane world volume directions contain $x^8, x^9$, 
which are the transverse directions of D7 brane. For example we can take the 
world volume directions of D3 are $x^6, x^7, x^8, x^9$. The D3-D7 
sectors arising from this geometry are T-dual to D1-D9 string 
configurations if $x^8, x^9$ are coordinates of $T^2$, 
which is S-dual to the heterotic string. On the 
intersection of D7 worldvolume and D3 world volume is the worldsheet 
of D1 brane. This configuration is closely related to the D1
instanton of Type I string theory. 
We will see that these two configurations will arise for the IIB 
orientifold on $K3 \times T^2/Z_2$, examples we consider later. 

Let's consider the first possibility. 
In the Type IIB theory, we have two chiral 10-d spinors $\epsilon_L,
\epsilon_R$
satisfying $\Gamma_{11}\epsilon_L=\epsilon_L,
\Gamma_{11}\epsilon_R=\epsilon_R$.
In the presence of D7 branes, the unbroken supersymmetries are given by 
\begin{equation}
\epsilon_L=\Gamma_0\Gamma_1 \cdots \Gamma_7
\epsilon_R=\Gamma_8\Gamma_9 \epsilon_R
\end{equation}
 while in the presence of the D3 branes, we have further constraint 
\begin{equation}
\epsilon_L=\Gamma_4\Gamma_5\Gamma_6\Gamma_7 \epsilon_R.
  \label{susy1}
\end{equation}
From these two equations, we obtain the condition 
\begin{equation}
\epsilon_R =\Gamma_{456789} \epsilon_R.   \label{chiral}
\end{equation} 
Strictly speaking, we can impose the two conditions (\ref{susy1}), (\ref{chiral})
only if the positions of D3 branes and D7 branes along $x^8, x^9$ directions coincide.
If the D7 branes are away from the D3 branes, surviving supersymmetries on the D3 branes
are given by (\ref{susy1}). However in the evaluation of the superpotentials using 
the physical gauge approach, we use the Kaluza-Klein approximation so that compactified 
space is small. In this case, the zero modes of the spinors should satisfy (\ref{susy1}),
(\ref{chiral}) simultaneously. This could be understood better if we consider 
$x^8, x^9$ directions are compactified on a torus $T^2$. This is T-dual to the D5-D9 
configuration, where the D5 branes have 8 component surviving supersymmetries. The small 
radius limit of $T^2$ of the D3-D7 configurations corresponds to the large radius 
limit of D5-D9 configurations. 
With the conditions (\ref{susy1}), (\ref{chiral}) satisfied, 
the decomposition of the 10-d chiral spinor is given by 
\begin{eqnarray}
S_{10}^{+} & \equiv & (S_6^+ \otimes S_4^+) \oplus (S_6^-\otimes S_4^-)  \nonumber \\
S_6^+  & \equiv & (S_D^+\otimes S_N^+) \oplus (S_D^-\otimes S_N^-)
\end{eqnarray}
where $S_6^+, S_D^+, S_N^+$ are a positive chirality spinor in the Calabi-Yau
manifold, in the 4 cycle and in the normal direction to the 4-cycle within
the Calabi-Yau manifold respectively while $S_6^-, S_D^-, S_N^-$ are
negative chirality spinors defined on the corresponding spaces.
Among 16 components of $S_{10}^+$ only 8 components  $S_6^+\otimes S_4^+$ are 
consistent with the above supersymmetry (\ref{susy1}), (\ref{chiral}). 
In order to proceed 
with the spacetime approach of D-brane instantons, one needs a $\kappa$ invariant 
action of D3 brane in the presence of D7 branes. This is not constructed yet. 
However we just need the quadratic part of the D3 brane action for that purpose
and one important ingredient is the structure of worldvolume fermions. 
The $\kappa$ invariant action is written in terms of 10-d spacetime spinor.
Upon the static gauge fixing, this turns into worldvolume spinors of D3 brane 
and the unbroken supersymmetries are given by (\ref{susy1}) if the D3 brane is separated 
from the D7 branes along $x^8, x^9$ directions while those are given by (\ref{susy1})
and (\ref{chiral}) if the D3 brane is coincident with the D7 branes. 
Upon Kaluza-Klein reduction, the zero modes of the spinors satisfy (\ref{susy1}) and 
(\ref{chiral}) simultaneously. We assume that such $\kappa$ invariant action can be 
constructed. 
This is also consistent with the low energy supersymmetric theory obtained from the D3-D7 
configurations, which is the theory of 8 supercharges with bifundamentals coming from
D3-D7 sectors. 
Related example of  $\kappa$ invariant action of membrane of M-theory with boundaries 
was constructed by \cite{Cederwall} where in the bulk we have 32 supersymmetries before 
the static gauge fixing while in the boundaries we have 16 supersymmetries. Upon the 
double dimensional reduction we obtain the $\kappa$ invariant action of the heterotic 
string with 16 supersymmetries before the gauge fixing \cite{Park1}.
This is S-dual to D1-D9 configuration which is T-dual to D3-D7 configuration 
considered later. In this example, one can check explicitly that the above line of argument 
is correct. With the above arguments,
we regard $S_6^+ \otimes S_4^+$ as D3 brane worldvolume spinors. 

Using the identification between spinors and forms on a Kahler manifold\cite{Witten2},
$S_D^{\pm}$ is decomposed as 
\begin{eqnarray}
S_D^+ &\simeq & K_D^{\frac{1}{2}}\otimes(\Omega^{(0,0)}\oplus\Omega^{(0,2)}) \nonumber\\
S_D^- &\simeq & K_D^{\frac{1}{2}}\otimes \Omega^{(0,1)}
 \label{spinorform2}
\end{eqnarray}
where $K_D$ is the canonical bundle of the four-cycle $D$. Using the adjunction formula 
for the normal bundle $N$ in a Calabi-Yau manifold, $N=K_D$ we have 
\begin{equation}
S_N^+\simeq K_D^{\frac{1}{2}}, \,\,\, S_N^- \simeq K_D^{-\frac{1}{2}}
\end{equation}
so that 
\begin{eqnarray}
S_D^+ \otimes S_N^+ &\simeq & (K_D^{\frac{1}{2}}\otimes
(\Omega^{(0,0)}\oplus\Omega^{(0,2)}))\otimes K^{\frac{1}{2}}_{D\frac{1}{2}}  \\
  &\simeq & K_D\otimes \Omega^{(0,0)} \oplus K_D\otimes \Omega^{(0,2)} \nonumber  \\
S_D^- \otimes S_N^- &\simeq &  (K_D^{\frac{1}{2}}\otimes 
\Omega^{(0,1)})\otimes K^{-\frac{1}{2}}_{D-\frac{1}{2}}
=\Omega^{(0,1)} \nonumber
\end{eqnarray}
here we denote the $U(1)$ charge of the $S_N^{\pm}$ as a subscript where  $U(1)$
is the rotation group of the normal direction to the 4-cycle. 
We have 
\begin{eqnarray}
S_6^+ & \simeq & (K_D\otimes \Omega^{(0,0)})\oplus (K_D\otimes \Omega^{(0,2)})
\oplus \Omega^{(0,1)}  \nonumber \\
S_6^- & \simeq &\Omega^{(0,0)}\oplus \Omega^{(0,2)}
\oplus (K_D\otimes\Omega^{(0,1)}) \label{spinorform} 
\end{eqnarray}
We introduce the complex coordinates $z^a, z^{\bar{a}}$ for the
4-cycle
and $z, \bar{z}$ for the normal direction to the 4-cycle in the
Calabi-Yau manifold.
The Dirac equation $D: S_6^+ \rightarrow S_6^-$ is of the form 
\begin{equation}
(\gamma^a
\nabla_a+\gamma^{\bar{a}}\nabla_{\bar{a}})\theta=0
\end{equation}
We will use the standard arguments about spinors on Kahler manifolds 
to make the identification between spinors and forms explicit\cite{Kallosh}. 
To this end, we let the gamma matrices act as 
\begin{equation}
\Gamma^{\bar{a}}=dz^{\bar{a}}\wedge, \,\,\,
\Gamma^{a}=g^{a\bar{b}}i_{\bar{b}}
\end{equation}
where $i_{\bar{b}}$ denotes a contraction on the differential forms 
and we define a Clifford vacuum as a state satisfying 
\begin{equation}
\Gamma^z|\Omega>=0,  \,\,\, \Gamma^a|\Omega>=0.
\end{equation}
In this formalism   elements of $S_6^+$ can be written as 
\begin{equation}
(\phi_{\bar{z}}\Gamma^{\bar{z}}+\phi_{\bar{z}\bar{c}\bar{d}}\Gamma^{\bar{z}\bar{c}\bar{d}}
+\phi_{\bar{a}}\Gamma^{\bar{a}}) |\Omega>
\end{equation}
The resulting Dirac equation is 
\begin{eqnarray}
\partial^{\bar{b}}\phi_{\bar{b}} &=& 0 \nonumber \\
\partial_{[\bar{a}}\phi_{\bar{b}]} &=& 0 \nonumber\\
\partial_{\bar{a}}^A\phi_{\bar{z}}+2\partial^{\bar{b}A}\phi_{\bar{z}\bar{b}\bar{a}}
&=& 0
\end{eqnarray}
On forms which has $\bar{z}$ index, we use a covariant derivative 
$\partial^A\equiv \partial+A$ rather than the usual derivative. 
Without the presence of the flux, the forms appearing in the above
expression are harmonic so the number of zero modes 
of the Dirac operator $D: S_6^+ \rightarrow S_6^-$ is given by the arithmetic genus
$\chi_D \equiv h^{0,0}-h^{0,1}+h^{0,2}$ where the zero modes are counted as positive
for the positive chirality spinor on the D3 brane and as negative for
the negative chirality spinor. Since
$S_4^+$ is of rank two, if the index $D$ is equal to one, 
we have two fermion zero modes, which is the
condition
required for the nonzero superpotential contribution. Note that this
counting is related to $U(1)$ anomaly of the rotation in the
transverse direction where $U(1)$ charge of $S_2^+$ is $1/2$ while
that of $S_2^-$ is $-1/2$. The $U(1)$ anomaly is the same as $\chi_D$. 
The anomaly of the $U(1)$ symmetry plays the crucial role in Witten's work on the 
nonperturbative superpotentials due to M5 branes\cite{Witten2}. 
The above mode counting is done on the Calabi-Yau manifold before taking orientifolding
action. If we consider
the orientifold we should consider the further projection due to the 
orientifold to make it sure that the zero modes 
are surviving from the
orientifold projections. We take the procedure to consider the cycles 
of the Calabi-Yau manifold before the orientifold action and consider
the further restrictions coming from the orientifolding action. The simple case 
would be to consider D3 brane on  four-cycles which are not fixed by the 
orientifold actions. 
It is argued in \cite{Gorlich}, if M5-brane wrapping a cycle $D\rightarrow S$
which are fibrations over $S$ with $P^1$ fiber, the arithmetic genus of $D$
\begin{equation}
\chi(D)=\Sigma_{1=0}^{3} (-1)^ih^{0,i}(D)=h^{0,0}(S)-h^{0,1}(S)+h^{0,2}(S)
\end{equation}
using the Leray spectral sequence. Since we expect M5 brane wrapping on $P^1$ is mapped 
to D3 brane, M5-brane zero mode counting is consistent with that of D3 brane.
\footnote{It's known that M5 brane wrapping on $K3$ is dual to heterotic string 
or to Type I D1 string\cite{Cherkis}. If we consider the elliptic $K3$ and if we T-dualize 
fiberwise along the elliptic fiber, we obtain D3 brane in the dual side. 
For generic fiber, this holds. Bad fibers of elliptic fibration give rise to 
the D3-D7 sectors.}

The above zero mode analysis is carried out for a single D3 brane
wrapping on a 4-cycle in a Calabi-Yau manifold. Here we make an
important assumption that there are no massless D3-D7 sectors. 
However at the special point of moduli space we can have massless
D3-D7 sectors. 
Once the Kaluza-Klein reduction is
carried out, D7 branes and D3 branes are turning into D3 branes and
D(-1) branes respectively from the 3+1 dimensional point of view. 
Thus D3-D7 brane configuration is reduced to the small instanton
configuration along $0,1 ,2 ,3$ directions. Thus we have to consider
a small instanton in $R^4$ with the assumption that the usual Kaluza-Klein 
reduction being valid. The additional moduli space we should
integrate over is just the instanton moduli space of the super
Yang-Mills theory obtained from the dimensional reduction of D7 branes wrapping
the 4-cycle. Here the gauge coupling is related to the volume of
the 4-cycle. 
In \cite{Witten3}, the dimension of the instanton moduli space is
counted as the independent of the number of hypermultiplets of D5
brane
in the presence of D9 branes, which are T-dual to our D3-D7
configurations.
Let us denote the gauge group arising from D9 branes by $SO(N)$ and
that from D5 branes by $Sp(k)$ then we have one hypermultiplet
transforming $(N, 2k)$ arising from D5-D9 sectors while we have one
hypermultiplet transforming as antisymmetric representation of $Sp(k)$
arising 
from D5-D5 sectors. When $k=1$ this antisymmetric representation is
nothing but the ordinary scalars representing the fluctuations of the 
D5-brane position. The dimension of the instanton moduli space is
given by 
\begin{equation}
4Nk+4\frac{2k(2k-1)}{2}-4\frac{2k(2k+1)}{2} =4Nk-8k=4k(N-2)
\end{equation}
Related to the fermion zero mode counting, the important thing is that
we have additional D-term constraints whose number is the same as the
adjoint of the D5 brane gauge group. 
When we consider the case with D9 brane gauge group being $U(N)$ and
D5 brane gauge group being $U(k)$ the dimension of the instanton
moduli space is given by $4Nk+4k^2-4k^2=4Nk$ where the first factor is 
the contribution from the D5-D9 sectors, the second factor is that
from the D5-D5 sectors while the last one comes from the D-term
constraints. When we consider the special case of D9-brane gauge group
being $U(1)$, the dimension of the moduli space is $4k$ where the $4k$
can be regarded as the position of $k$ D5 branes transverse to its
worldvolume but along the D9 brane world volume direction. Translated
to D3-D7 configurations of our interest, these $4k$ hypermultiplets 
denote the position of $D3$ branes along $0,1,2,3$ directions. 

Now we should consider the fermion zero modes. According to the field
theory result, the fermion zero modes are $4Nk$ if we have $N=2$
supersymmetry
in the four-dimensions, while those are $2Nk$ if we have $N=1$ supersymmetry for pure 
supersymmetric gauge theory without matter\cite{Mattis}. 
Especially for $U(1)\times U(1)$, i.e., the abelian D7-brane
gauge group and a single D3 brane wrapping around a 4-cycle, we have 
2 fermion zero modes, which is needed for a nonzero superpotential. 
Here we see that we can have the nontrivial superpotential from an 
abelian instanton configuration. Note that while two fermion zero modes 
follow from the previous analysis of D3-D3 sectors if D3-D7 sectors are massive,
the counting of the zero modes with the massless D3-D7 sectors 
is the result of the D3-D3 sectors and D3-D7 sectors 
combined with the D-term constraints. Also note that in this configuration the analysis 
of the multiple D3 instantons is reduced to the multi instanton calculus 
in the field theory. This would be an interesting topic to pursue further. 
Note that if we have $SU(N)$ gauge group from the D7-branes and a
single $D3$ brane the number of fermion zero mode is $2N$, which
agrees with the number of fermion zero modes of one instanton for $SU(N)$ supersymmetric
gauge theory. 

If we consider a simple case of a rigid 4-cycle with $h^{(1,0)}=0$ within the Calabi-Yau, then 
we just have 4 translational bosonic zero modes along $x^0$ to $x^3$ directions
and 2 fermion zero modes, which are the Goldstone fermion zero modes associated 
with the breaking of the supersymmetry in the presence of the D3 instanton.
The superpotential expression is given by\cite{Witten, Park1}
\begin{equation}
W={\rm exp} (-A +i\int_D C) \frac{{\rm Pffaf}' D_F}{\sqrt{\rm det}' D_B}
\end{equation}
if the D3-D7 sectors are massive while the integration of the instanton 
moduli space should be considered if D3-D7 sectors are massless. 
Here $A$ denotes the volume of the 4-cycle $D$ and $C$ is a RR 4-form potential,
the prime means that we are considering the determinant factors 
for nonzero modes and ${\sqrt{{\rm det}' D_B}}$ comes from the one-loop determinant 
of the bosonic modes and ${\rm Pffaf}' D_F$ comes from that of the fermionic modes
of the D3 branes. The exponential terms come from the classical instanton action, 
while the other factors represent the one-loop integral over the quantum fluctuations 
around the classical instanton solutions. The superpotential generated by 
the instanton, apart from the usual exponential terms,  
is independent of the Kahler class of the Calabi-Yau manifold $M$ 
and so can be computed by scaling up the metric of $M$. 
The one-loop determinant is invariant under the scaling while higher loop 
corrections to the worldvolume computation would be proportional to the inverse 
power of the Kahler class and so vanish by holomorphy. The one-loop approximation 
to the superpotential is exact \cite{Witten2}. 

Let's now turn to the second possibility. 
We assume that D7 brane worldvolume direction spans $x^0$ to $x^7$
while D3 brane worldvolume spans $x^4,x^5, x^8, x^9$ directions. 
From D7 branes we have the unbroken supersymmetry
\begin{equation}
\epsilon_L=\Gamma_8\Gamma_9 \epsilon_R
\end{equation}
 while from the D3 brane 
we have the condition
\begin{equation}
\epsilon_L=\Gamma_{0123}\Gamma_{67}\epsilon_R=\Gamma_{4589}\epsilon_R.
\end{equation}
And from these we obtain $\epsilon_L=\Gamma_{45}\epsilon_L$.
Note that along the intersection of D7 and D3 we have a two-dimensional 
worldsheet and the supersymmetry above represents the surviving
supersymmetry on this two-dimensional worldsheet. Note that this represents a chiral
theory in two dimensions. This is consistent with the fact that upon the
T-dualities the D7-D3 configurations are mapped to D1-D9
configurations, which is D1-string configuration in Type I theory, S-dual to the
heterotic string. Since the surviving supersymmetry is different from 
the previous one, the zero mode analysis is also different. 
Again we decompose the spinors
$S_{10}^+=(S_6^+ \otimes S_4^+)\oplus (S_6^- \otimes S_4^-)$.
Now let $N$ be the normal direction of D3 brane so that 
\begin{eqnarray}
S_6^+ &\simeq & (S_N^+ \otimes S_D^+) \oplus  (S_N^- \otimes S_D^-)  \nonumber
\\
S_6^- &\simeq & (S_N^+ \otimes S_D^-) \oplus  (S_N^- \otimes S_D^+)
\end{eqnarray}
Now decompose the spinors on the D3 brane world-volume along the two-dimensional
worldsheet 
and its normal direction 
\begin{eqnarray}
S_D^+ &\simeq & (S_2^+\otimes S_{\tilde{N}^*}^+) \oplus  (S_2^-\otimes
S_{\tilde{N}^*}^-)  \nonumber  \\
S_D^- &\simeq & (S_2^+\otimes S_{\tilde{N}^*}^-) \oplus  (S_2^-\otimes
S_{\tilde{N}^*}^+)  
\end{eqnarray}
where we define the spinors $S_{\tilde{N}^*}$ associated with the
holomorphic conormal bundle to the two-dimensional worldsheet $C$ so that 
$S_{\tilde{N}^*}^+\sim (\tilde{N}^*)^{\frac{1}{2}}, \,\, 
S_{\tilde{N}^*}^-\sim (\tilde{N}^*)^{-\frac{1}{2}}$ and $S_2$ be the
spinors associated with the canonical bundle on the two-dimensional worldsheet
so that $S_2^+\sim K_2^{\frac{1}{2}} , \,\, S_2^-\sim
K_2^{-\frac{1}{2}}$.
This convention is consistent with the identification of the spinors
with the antiholomorphic forms tensored with the squareroot of the
canonical bundle, eq. (\ref{spinorform2}). Note that 
\begin{equation}
S_2^+\oplus S_2^- \simeq K_2^{\frac{1}{2}}\oplus K_2^{-\frac{1}{2}}
\simeq  K_2^{\frac{1}{2}}\otimes (\Omega^{(0,0)}(C)\oplus \Omega^{(0,1)}(C))
\end{equation}
 and
similarly
\begin{eqnarray}
S_D &\simeq & (S_2^+\oplus S_2^-)\otimes (S_{\tilde{N}^*}^+\oplus 
S_{\tilde{N}^*}^-)  \nonumber \\
&\simeq &  (K_2^{\frac{1}{2}}\oplus K_2^{-\frac{1}{2}})\otimes 
(\tilde{N}^{*\frac{1}{2}} \oplus \tilde{N}^{*-\frac{1}{2}})  \nonumber
\\
&\simeq & K_2^{\frac{1}{2}}\otimes \tilde{N}^{*\frac{1}{2}} (1\oplus
K_2^{-1}\oplus\tilde{N}^{*-1} \otimes (K_2\otimes \tilde{N}^*)^{-1})  \nonumber
\\
&\simeq & K_D^{\frac{1}{2}}\otimes(\Omega^{(0,0)}(D)\oplus \Omega^{(0,1)}(D)\oplus
\Omega^{(0,2)}(D))
\end{eqnarray}

If we decompose the 16 component spinors $S_{10}^+$ in terms of $S_2,
S_{\tilde{N}^*}, S_{N}$ the surviving 8 components are given by 
\begin{eqnarray}
S_N^+\otimes S_2^+\otimes S_{\tilde{N}*}^+ \otimes S_4^+ &\subset & S_6^+
\otimes S_4^+  \nonumber \\
S_N^-\otimes S_2^+\otimes S_{\tilde{N}*}^- \otimes S_4^+ &\subset & S_6^+
\otimes S_4^+  \nonumber \\
S_N^+\otimes S_2^+\otimes S_{\tilde{N}*}^- \otimes S_4^- &\subset & S_6^-
\otimes S_4^-  \nonumber \\
S_N^-\otimes S_2^+\otimes S_{\tilde{N}*}^+ \otimes S_4^- &\subset & S_6^-
\otimes S_4^-  \label{spinor}
\end{eqnarray}

Depending on the embedded geometry of the 2d worldsheet in the Calabi-Yau
one can have different number of zero modes. 
If we choose the 2d worldsheet to be a rigid $P^1$, then the normal bundle
over $P^1$ is described by $O(-1)\oplus O(-1)$ bundles over $P^1$, where
$O(n)$ is a holomorphic line bundle whose sections are functions 
homogeneous of degree $n$ in the homogeneous coordinates of $P^1$.
Then $S_2^+\simeq O(-1), S_2^-\simeq O(1), S_{\tilde{N}*}^+\simeq O(1/2), 
S_{\tilde{N}*}^-\simeq O(-1/2)$ and $S_N^+\simeq O(-1/2), S_N^-\simeq
O(1/2)$\footnote{Since we are considering the tensor products of the spinors, 
all of the relevant expressions are well defined}
one can see that from eq. (\ref{spinor}) only 
\begin{equation}
S_N^-\otimes S_2^+\otimes S_{\tilde{N}*}^+ \otimes S_4^- \sim
O\oplus O
\end{equation}
 contributes to the zero modes and 
$S_N^-\otimes S_2^+\otimes S_{\tilde{N}*}^+ \simeq \Omega^{(0,0)}(D)$ if we
use the identification eq. (\ref{spinorform}). 

 The D3-D7 sector 
represents the chiral current algebra of $SO(32)$ and can be represented 
as left-moving fermions. The action is given by\cite{Park1}
\begin{equation}
S_L=\int_C d^2 \sigma \sqrt{g} \bar{\Psi}^a \gamma^i (D_i \delta^{ab}-A_i^{ab}) \Psi^b
\end{equation}
where $a, b$ denote $SO(32)$ indices and $A_i=A_\mu 
\frac{\partial X^{\mu}}{\partial \sigma^i}$ is a pullback of the spacetime 
field $A_{\mu}$. If we consider the D1-D9 configurations, $A_{\mu}$ are simply
$SO(32)$ bundle configuration. In the case of D3-D7 configurations
$A_8, A_9$ are position moduli of D7 branes while the other components 
represent bundle configurations. Upon pullback to the worldsheet, the combination 
of position moduli and the bundle moduli induces a background gauge field 
on the world-sheet. Let $V$ denote such induced field. 
Let us denote the left-handed spin bundles of 
the 2d-worldsheet $C$ by $S_-$. In a suitable complex structure the kinetic 
operator for a left moving fermion is a $\bar{\partial}$ operator\cite{Witten}. 
The left moving 
fermions are a section of $S_-\otimes V$. 
The superpotential expression for the D3 brane wrapping a rigid cycle is 
\begin{equation}
W={\rm exp} (-A +i\int_D C) \frac{{\rm Pffaf}' D_F}{\sqrt{{\rm det}' D_B}} 
{\rm Pffaf}(\bar{\partial}_{S_-\otimes V})
\end{equation}
If we consider the the 2-d worldsheet of D3-D7 sectors is $P^1$, it is
analyzed that the fermion determinant is nonzero only if the bundle
restricted on $P^1$ is trivial. 
Thus in addition to the usual zero mode analysis
concerning on the D3-D3 sectors, we should check D3-D7 sectors give
rise to nontrivial fermion determinant. This could be a severe
restriction for the generic $SO(32)$ bundle configurations. 
Such nontrivial examples were worked out in the heterotic M-theory
setting in \cite{Ovrut}. 
In a simple example of $K3 \times T^2$, the D9 brane configurations are 
simply $SO(32)$ instanton bundles along $K3$ and Wilson lines along $T^2$. 
T-dualizing to D7 branes wrapping on $K3$, we have a collection of D7 branes located 
at various points on $T^2$ with instanton bundles along $K3$. 
For each D7 brane group located at a separate point on $T^2$ we have the pullback of 
bundles of $K3$ into the two-dimensional worldsheet. The Pffafian on the 
worldsheet is the product of all such contributions. 
This Pffafian factor gives rise to the additional dependence of the superpotential 
on the vector bundle moduli in the Type I theory\cite{Ovrut}. Translated to our case, 
this implies the 
dependence of the superpotential on the position and shape moduli of D7 branes 
as well as the bundle moduli on the D7 branes. It would be interesting to find 
nontrivial examples where explicit dependences could be exhibited. 

\section{D3 instanton effects in the presence of the flux}

Now we consider the D3 instanton effects for a Calabi-Yau orientifold 
in the presence of the fluxes. 
In \cite{Trivedi}, fermion mass term is derived in the presence of the flux
using the D3-brane action with $\kappa$ symmetry and taking the static
gauge consistent with our cases. This is needed in the evaluation of the one loop determinant
appearing in the superpotential expression. D-brane actions in a general bosonic 
backgrounds in component form are considered by \cite{Martucci}, for example. 
The result is that for the D3 brane wrapping a four-cycle, the quadratic
action $S_2=S_k+S_{mass}$ with
\begin{eqnarray}
S_k &=& -\mu_3\int d^4 x \sqrt{det g} 
(\frac{1}{2}e^{-\phi}\bar{\Theta}\Gamma_k D^{\bar{k}}\Theta)
\nonumber \\
S_{mass} &=& -\mu_3\int d^4 x \sqrt{det g} \bar{\Theta}
(e^{-\phi}\frac{1}{48}\Gamma^{mnp}H_{mnp}-\frac{1}{16}e^{-\phi}\Gamma_{\bar{i}pq}H^{\bar{i}pq}
\nonumber \\
 & &
 -\frac{1}{32}\epsilon^{\bar{i}\bar{j}\bar{k}\bar{l}}\Gamma^p_{\,\, \bar{i}\bar{j}}
F'_{\bar{k}\bar{l}p})\Theta
\end{eqnarray}
Here we assume that the fermion mass terms arise due to the 3-form
fluxes and there are no $F-B$ terms in the D3 brane worldvolume. 
And $H$ denotes NS-NS 3-form, $F'$ denotes RR 3-form and $\bar{i},
\bar{j}, \bar{k}, \bar{l}$ are along the worldvolume while $m,n,p$ takes 
0 to 9 in spacetime. It's easy to see that if $H, F'$ have two legs
along the brane and one leg along the normal direction then 
\begin{eqnarray}
S_{mass} &= &\mu_3\int d^4 x \sqrt{det g} \bar{\Theta}
(-\frac{1}{16}e^{-\phi}\Gamma_{\bar{i}\bar{j}q}H^{\bar{i}\bar{j}q}-
\frac{1}{32}\epsilon^{\bar{i}\bar{j}\bar{k}\bar{l}}\Gamma^p_{\,\, \bar{i}\bar{j}}
F'_{\bar{k}\bar{l}p})\Theta  \nonumber \\
 &=& \mu_3\int d^4 x \sqrt{det g} \bar{\Theta}
(-\frac{1}{16}\Gamma_{\bar{i}\bar{j}q}(e^{-\phi}H^{\bar{i}\bar{j}q}
+(*F')^{\bar{i}\bar{j}q})\Theta  \label{massterm}
\end{eqnarray}
where the Hodge star is defined for the D3 worldvolume. 
The terms appearing in eq. (\ref{massterm}) breaks the U(1) symmetry in the 
normal direction 
allowing in particular two fermions with opposite sign charge to pair up
and get heavy, which suggests that the zero mode analysis could be different
in the presence of the fluxes.
We define $G\equiv F'-ie^{\phi}H$ for later purposes.

As an example of the Calabi-Yau orientifold with the fluxes, we consider 
a simple example, 
IIB orientifold on $K3 \times T^2/Z_2$.
According to Sen\cite{Sen}, F-theory on $K3$ is equivalent to IIB orientifold on
$T^2/Z_2$ with the orientifold action $\Omega R_{89} (-1)^{F_L}$. Thus 
IIB orientifold on $K3 \times T^2/Z_2$ is dual to F-theory on $K3
\times K3$. Closely related theory, the M5 brane instanton on M
theory on $K3 \times K3$ was discussed in \cite{Kallosh, Saulina}. 
Here we consider a simple orientifold which is dual to F-theory on $K3
\times K3$ with flux $G_4=\Omega_1 \wedge \bar{\Omega}_2 +
\bar{\Omega}_1\wedge \Omega_2$ where 
$\Omega_1, \Omega_2$ are holomorphic two forms of $K3$s . 
This is known to have N=1 supersymmetry\cite{Gorlich, Trivedi2}. 
The relation between the 4-form in F-theory is given by
\begin{equation}
G_4=-\frac{1}{\phi-\bar{\phi}}G_3 \wedge d\bar{z'}
+\frac{1}{\phi-\bar{\phi}}\bar{G_3} \wedge dz'
\end{equation}
where 
$z'$ is the elliptic fiber direction of the F-theory. Thus $G_3$ on the
orientifold $K3 \times T^2/Z_2$ is given by 
\begin{equation}
\Omega \wedge d\bar{z}
\end{equation}
where $\Omega$ is the holomorphic two-form on $K3$ and $z$ is a holomorphic 
coordinate of $T^2$. 
In a local coordinates $G_3$ has a nontrivial component
$G_{ab\bar{z}}$ and $G_{\bar{a}\bar{b}z}$ where $z$ is a local holomorphic 
coordinate on $T^2$. For simplicity we consider
the case with the constant dilaton. This occurs if the tadpole
cancellation occurs locally. The resulting gauge group is $SO(8)^4$. 
If we consider the D3
instantons, there are two types of four-cycles we can consider.
The fist one is the D3-brane wrapping on $K3$. And the second case 
is D3-brane wrapping on $P^1 \times T^2/Z_2$ where $P^1$ is a holomorphic 
curve in $K3$.

\subsection{First case: D3 brane wrapping on $K3$} 
We introduce the complex coordinates $z^a, z^{\bar{a}}$ for the
$K3$
and $z, \bar{z}$ for the normal directions to $K3$ in the
Calabi-Yau manifold.
The Dirac equation $D: S_6^+ \rightarrow S_6^-$ is of the form 
\begin{equation}
(\gamma^a
\nabla_a+\gamma^{\bar{a}}\nabla_{\bar{a}}+G_{ab\bar{z}}\Gamma^{ab\bar{z}}
+G_{\bar{a}\bar{b}z}\Gamma^{\bar{a}\bar{b}z})\theta=0
\end{equation}
after absorbing constants into the definition of $G$ in eq.(\ref{massterm}). 
With the same identification between spinors and forms,
the resulting Dirac equation is 
\begin{eqnarray}
\partial^{\bar{b}}\phi_{\bar{b}} &=& 0  \\
\partial_{[\bar{a}}\phi_{\bar{b}]}
+G_{[\bar{a}\bar{b}]z}\phi^z &=& 0 \label{eqf}\\
\partial_{\bar{a}}^A\phi_{\bar{z}}+2\partial^{\bar{b}A}\phi_{\bar{z}\bar{b}\bar{a}}
&=& 0
\end{eqnarray}

Without the presence of the flux, the index $D$ on the D3 brane wrapping on $K3$
is given by $h^{0,0}-h^{0,1}+h^{0,2}=2$, which would not contribute to the 
superpotential. However the presence of the flux changes the zero mode analysis.
Here we can use the similar trick to \cite{Kallosh}.
We introduce the projector $H$ onto harmonic forms so that $H(\omega)$
is a harmonic form (possibly zero) for any form $\omega$. The
projector gives zero on any exact or co-exact form
\begin{equation}
H(\bar{\partial}\omega)=0,  \,\,\,
H(\bar{\partial}^{\dagger}\omega)=0  \,\,\, \forall \omega .
\end{equation}
By acting $H$ on eq. (\ref{eqf}) we have 
\begin{equation}
H(G_{\bar{a}\bar{b}z} \phi^z dz^{\bar{a}} \wedge dz^{\bar{b}})=0. \label{projection}
\end{equation}
These are $h^{2,0} $ equations for $h^{2,0}$ variables.
One expects that generically $\phi^z$ satisfying this condition is
trivial. 
For a nonzero mode, one uses the formula $1-H=\Delta G$ where $\Delta$
is the Laplacian and $G$ is the Green function of the corresponding
Laplacian
\begin{equation}
(1-H)(\omega)=(\bar{\partial}\bar{\partial}^{\dagger}
+\bar{\partial}^{\dagger}\bar{\partial})G\omega  \,\,\,\forall \omega
\end{equation}
to obtain from the eq. (\ref{eqf})
\begin{eqnarray}
(1-H)G_{\bar{a}\bar{b}z} \phi^z dz^{\bar{a}} \wedge dz^{\bar{b}}
&=&G_{\bar{a}\bar{b}z} \phi^z dz^{\bar{a}} \wedge dz^{\bar{b}}  \\
&= &(\bar{\partial}\bar{\partial}^{\dagger}
+\bar{\partial}^{\dagger}\bar{\partial})G 
G_{\bar{a}\bar{b}z} \phi^z dz^{\bar{a}} \wedge dz^{\bar{b}}    \\
&= & -\partial_{\bar{a}}\phi_{\bar{b}}dz^{\bar{a}} \wedge dz^{\bar{b}}
\end{eqnarray}
From this we obtain one special solution
\begin{equation}
\phi_{\bar{b}0}=-2g^{\bar{a}a}\partial_a (G G_{\bar{a}\bar{b}z}\phi^z).
\end{equation}
Now one can add $h^{(1,0)}$ zero modes to this special solution. Since 
the equation governing $h^{(0,0)}$ is not changed, 
 the number of zero modes are given by
$h^{(0,0)}-h^{(1,0)}+n$ with 
$n$ being the dimension of the solution space satisfying (\ref{projection}).
In particular, if $G=\Omega\wedge d\bar{z}$, since 
$\phi^z=g^{z\bar{z}}\phi_{\bar{z}}$ is an element of 
$K\otimes \Omega^{(0,0)}\equiv\Omega^{(2,0)}$ this is just an element of 
1-dimensional vector space. Multiplied by $\Omega$, 
$G_{\bar{a}\bar{b}z} \phi^z dz^{\bar{a}} \wedge
dz^{\bar{b}}$ is proportional to the contraction of $\Omega$ and $\bar{\Omega}$.
Thus $H(G_{\bar{a}\bar{b}z} \phi^z dz^{\bar{a}} \wedge
dz^{\bar{b}})$ is nonzero unless $\phi^z$ is nonzero. Hence the 
number of zero modes 
$h^{(0,0)}-h^{(1,0)}+n=1$ so that this could contribute to the superpotential.
However we should make it sure that additional D3-D7 sector contribution 
does not give the null result. In order for a single D3-brane to contribute to 
the superpotential, we should have either massive D3-D7 sectors or there should be 
an abelian configuration of D7. In the spacetime approach of the evaluation of 
the superpotential, the D3 brane configuration also satisfies the classical 
equation of motion\cite{Witten}. 
In the presence of the flux, D3 brane feels the same potential 
as the D7 brane along the Calabi-Yau manifold. Thus the D3 brane could be located 
only at the minimum of the potential. It looks rather difficult to analyze the 
abelian D7 brane configuration in the presence of the flux since this 
generally have the dilaton gradient. Anyway if such configuration exists, we can 
put a single D3 brane at the single D7 brane. This would give rise to the nontrivial 
superpotential. It would be worthwhile to look for such configurations. 
However in the case at hand, we know for a field theory point of view that there 
are nontrivial superpotentials. In the constant dilaton configuration with $SO(8)^4$
gauge group, the gauge theory on the four groups of D7 branes are just N=1 supersymmetric 
gauge theory without additional supersymmetric matter. In this case we know there are 
superpotential due to the gaugino condensation. \footnote{If four-dimensional
 theory is compactified on a circle 
$S^1$, the superpotential is due to the magnetic monopoles. The superpotential does 
not depend on the radius of the circle and we can take the decompactification limit.
If we realize the four-dimensional theory by D3 brane, 
the magnetic monopole in three-dimension is represented 
as D0 brane stretching between the D2 branes which are separated on a dual circle 
$\tilde{S}^1$ upon T-dualizing the D3 branes with Wilson lines.}

Note that in terms of geometric engineering, this provides an interesting test bed. 
Since the effect of the flux is to give the mass to the adjoint matter 
this system is closely related example as analyzed in \cite{Hollowood}. 
Turning things around, the field theory analysis gives the expression for 
the superpotential for the flux where the adjoint mass is proportional to the 
flux strength. Since the measures of the multi instanton moduli spaces 
for supersymmetric gauge theories were worked out\cite{Mattis, Mattis2}, 
this is a good starting point 
to work out the multi-instanton effect of D3 branes. Here the multi instantons are 
represented by nonabelian configurations of D3 branes. It would be interesting 
to explicitly work out these.

\subsection{Second case: D3 brane wrapping on $P^1 \times T^2/Z_2$}

In this case massless modes of D7-D3 sectors always exist. 
Without the flux, D3 brane wrapping on $P^1 \times T^2/Z_2$ is T-dual
to D1 brane wrapping on $P^1$ and $T^2$ is transverse to D1 brane. 
The transverse geometry to $P^1$ of D1 worldvolume is $O\oplus O(-2)$ in 
the large volume limit
where $O$ denotes the $T^2$ directions. 
Without the flux this geometry has additional zero modes than is
needed for nonzero superpotential. The same is true of D3 brane and
one should be careful in dealing with $Z_2$ action acting on $T^2$,
which is not the usual geometric action but the combined action of the 
orientifold action and the geometric action. Since $P^1 \times T^2$ is mapped to
itself
under the orientifold action, 
the gauge group of U(N) of D3 branes are
reduced to $SO(N)$. The worldsheet degrees of freedom remain the same
so the zero mode structure is the same as before the orientifolding.
Here we have $S_2^+\simeq O(-1), S_2^-\simeq O(1), 
S_{\tilde{N}*}^+\simeq O, 
S_{\tilde{N}*}^-\simeq O$ and $S_N^+\simeq O(-1), S_N^-\simeq
O(1)$. Among eq. (\ref{spinor}), 
\begin{eqnarray}
S_N^+\otimes S_2^+\otimes S_{\tilde{N}*}^- \otimes S_4^- \simeq
O\oplus O  &\subset &S_6^+\otimes S_4^+ \nonumber  \\
S_N^-\oplus S_2^+\otimes S_{\tilde{N}*}^+ \otimes S_4^- \simeq O\oplus
O&\subset & S_6^-\otimes S_4^- 
\end{eqnarray}
contribute to the zero modes. Without the presence of fluxes, this is
more than needed for the nonzero superpotential. However in the
presence of the flux some of the zero modes are removed so that it can
contribute to the superpotential. Since 
$S_N^+\otimes S_2^+\otimes S_{\tilde{N}^*}^- \subset S_6^+$ and is an element
of $\Omega^{(0,1)}(D)$ from the analysis of the Dirac equation 
$D: S_6^+\rightarrow S_6^-$ one will see that $h^{(0,1)}$ is removed in the
presence of the flux.
If we denote the normal direction by a complex coordinate $y$
then the components of $G_3$ and $\bar{G}_3$ in the local coordinates 
can be written as 
$G_{yb\bar{c}}$ and $\bar{G}_{\bar{y}\bar{b}c}$. Here $y,b$ span the K3
directions while $b,c$ span the D3 worldvolume and $c$ denotes the
$T^2/Z_2$ direction. 
As in the previous case, we represent the spinor by forms. 
The Dirac equation for the D3 worldvolume fermion in the presence
of the flux is given by 
\begin{eqnarray}
\partial^{\bar{b}}\phi_{\bar{b}} &=& 0  \\
\partial_{\bar{a}}^A\phi_{\bar{y}}+2\partial^{\bar{b}A}(\phi_{\bar{y}\bar{b}\bar{a}})
+\bar{G}_{\bar{y}\bar{a}c}\phi^c &=& 0 \\
\partial_{\bar{a}}\phi_{\bar{b}}+G_{yc\bar{a}}\phi^{yc}_{\,\,\,\,\, \bar{b}}-(a \leftrightarrow b)
&=& 0
\end{eqnarray}
Again adopt the Harmonic projector, then we have 
\begin{equation}
H(\bar{G}_{\bar{y}\bar{a}c}\phi^c d\bar{z}\wedge d\bar{a})=0  \label{eqf2}
\end{equation}
and 
\begin{equation}
H(G_{yc\bar{a}}\phi^{yc}_{\,\,\,\,\, \bar{b}})-(a \leftrightarrow b)=0
\end{equation}
The first equation (\ref{eqf2}) is the map from $h^{(0,1)}$ to $h^{(0,1)}$
so generically remove all modes of $h^{(0,1)}$. 
The second equation is a map from $\Omega^{(0,0)}$ to $\Omega^{(0,2)}$.
In the case of our interest, since $h^{(0,2)}=0$ this is satisfied automatically.
For the first equation $\phi^c=g^{c\bar{c}}\phi_{\bar{c}}$ is an element of one dimensional
vector space $\Omega^{(0,1)}$ and 
$\bar{G}_{\bar{y}\bar{a}c}=\bar{\Omega}_{\bar{y}\bar{a}}\wedge dz^c$ is proportional to 
a nontrivial element of $\Omega^{(1,0)}$. Thus $\bar{G}_{\bar{y}\bar{a}c}\phi^{c}$
is proportional to the contraction of the nontrivial elements of $\Omega^{(1,0)}$
and $\Omega^{(0,1)}$ if $\phi^c$ is nontrivial. Thus $\phi^c$ should vanish 
in order for the eq. (\ref{eqf2}) to be satisfied. 

 On the other hand one can check 
that $S_N^-\otimes S_2^+\otimes S_{\tilde{N}*}^+ \subset S_6 ^-$ is an element
of $\Omega^{(0,0)}$ and by the similar analysis of the Dirac equation
$D: S_6^- \rightarrow S_6^+$ one sees that $h^{(0,0)}$ is retained in
the presence of the flux. Thus we are are left with the two fermion
zero modes so that it can contribute to the superpotential.
For $SO(8)^4$ configuration, for each group of D7 branes there are no nontrivial 
bundles so that the pullback is also trivial. Hence the Pfaffian factor 
gives the nonzero contribution. According to \cite{Aspinwall}, the $K3$ manifold 
we consider is rather special one, so called attractive $K3$. The Picard number of this 
$K3$ is 20 so that it has 20 holomorphically embedded $P^1$s. In our orientifold 
we have 21 Kahler moduli. Since we also have 21 types of D3 instantons available and 
all of them are nonzero, we expect that all of the Kahler moduli would be fixed 
by the instanton effects as happened in M theory on $K3 \times K3$\cite{Aspinwall}.
One final remark should be added. The above zero mode 
analysis is carried out assuming a specific complex structures for $K3 \times T^2$. 
However in the presence of the fluxes, the actual complex structure is not 
necessarily the same complex structure as we assume. However the number of zero 
modes remains the same as we deform the complex structures. In the actual 
evaluation of the nonzero superpotential, different complex structures give
different nonzero values.

\section{D5 instanton in Type I theory}
If we consider D5 brane wrapping on a Calabi-Yau, we expect the instanton effect 
$e^{-V+ \cdots}$ where $V$ is the volume modulus of Calabi-Yau. 
We can proceed the analysis in the similar way to that of D3 instanton.
For Type I, the unbroken supersymmetry is given by 
\begin{equation}
\epsilon_L=\epsilon_R
\end{equation}
and in the presence of D5 branes along Calabi-Yau we have 
\begin{equation}
\epsilon_L=\Gamma_{456789}\epsilon_R.
\end{equation}
Thus we have 
\begin{equation}
\epsilon_L=\Gamma_{456789}\epsilon_L.
\end{equation}
On the D5 brane world volume are there 8 component of spinors $S_6^+\otimes S_4^+$.
Since $S_6^+$ is the positive chirality spinor on the Calabi-Yau, 
we can identify them as 
\begin{equation}
S_6^+\simeq \Omega^{0,0}\oplus \Omega^{0,2}
\end{equation}
since the canonical bundle of the Calabi-Yau is trivial. 
The number of zero modes of the Dirac operator $D: S_6^+ \rightarrow S_6^-$ 
are given by $h^{0,0}+h^{0,2}$. For a generic Calabi-Yau $h^{(0,0)}=1, h^{(0,2)}=0$
and combined with the two degrees of freedom $S_4^+$, we have two fermion zero 
modes from the D5-D5 sectors of a single D5-brane. Again we should consider the 
effect of the D5-D9 sectors. Upon the Kaluza-Klein reduction, this is again 
reduced to the small instanton configuration in the 4d theory. Following the same 
logic in the D3 brane analysis, if we have the $SO(32)$ bundle configuration 
which breaks $SO(32)$ into $U(1)$ so that we are left with an abelian gauge 
group, the number of zero modes coming from D5-D5 and D5-D9 sectors with D-term constraints 
are precisely two so that we can have non-zero superpotential. 
But since the Calabi-Yau volume plays the role of the inverse gauge coupling of D9 sectors 
compactified on a Calabi-Yau for a nonabelian gauge group we can have the superpotential 
for the Calabi- Yau 
volume moduli, which makes the gaugino condensation possible. This depends on the matter 
content of the four-dimensional theory realized 
in D9 brane configuration compactified on the Calabi-Yau.
For a generic Calabi-Yau manifold, $h^{(0,1)}=h^{(0,2)}=0$. 
D9 brane gives rise to pure supersymmetric gauge theory.

\vspace{1cm}

\section*{Acknowledgments}
 We thank Sandip Trivedi for the correspondence and the useful discussion. 
The work of J.P. is supported by
Korea Research  Foundation (KRF) Grant KRF-2004-042-C00023.


\begin{thebibliography}{99}

\bibitem{KKLT} S. Kachru, R. Kallosh, A. Linde and S. Trivedi,
``De Ditter vacua in string theory, `` 
Phys. Rev. {\bf D}68 (2003) 046005, hep-th/0301240.

\bibitem{GKP} S. Giddings, S. Kachru and J. Polchinski, ``Hierarchies 
from fluxes in string compactifications,'' Phys. Rev. {\bf D66} (2002) 106006,
hep-th/0105097.

\bibitem{Moore} J. Harvey and G. Moore, ``Superpotentials and membrane 
instantons,'' hep-th/9907026. 

\bibitem{Witten} E. Witten, ``Worldsheet instanton corrections via D-instantons,''
hep-th/9907041.

\bibitem{Saulina} G. Moore and G. Peradze and N. Saulina, ``Instabilities in 
heterotic M theory induced by open membrane instantons,'' 
Nucl. Phys. {\bf B607} (2001) 117, hep-th/0012104.

\bibitem{Park1} E. Lima, B. Ovrut, J. Park and R. Reinbacher, `` Nonperturbative
superpotentials from membrane instantons in heterotic M theory,'' 
Nucl. Phys. {\bf B614} (2001) 117, hep-th/0101049.

\bibitem{Park2} E. Lima, B. Ovrut and J. Park, `` Five-brane superpotentials
in heterotic M theory,'' Nucl. Phys. {\bf B626} (2002) 113, hep-th/0102046.

\bibitem{Gorlich} L. Gorlich, S. Kachru, P. Tripathy and S. Trivedi, ``Gaugino 
condensation and nonperturbative superpotentials in flux compactifications,''
hep-th/0407130.

\bibitem{Mayr} P. Berglund and P. Mayr, ``Nonperturbative superpotentials in F-theory
and string duality,'' hep-th/0504058. 

\bibitem{Saulina2} N. Saulina, ``Topological constraints on stabilized flux vacua,''
hep-th/0503125.

\bibitem{Trivedi} P. Tripathy and S. Trivedi, ``D3 brane action and fermion zero 
modes in presence of background flux,'' hep-th/0503072. 

\bibitem{Martucci} D. Marolf, L. Martucci and P. Silva, ``Actions and fermionic symmetries 
for D-branes in bosonic backgrounds,'' {\bf JHEP 0307} (2003) 019, hep-th/0306066;
D. Marolf, L. Martucci and P. Silva, ``Fermions, T duality and effective actions 
for D-branes in bosonic backgrounds,'' {\bf JHEP 0304} (2003) 051, hep-th/0303209.

\bibitem{Kallosh} R. Kallosh and A. Kashani-Poor and A. Tomasiello, ``Counting 
fermion zero modes on M5 with fluxes,'' hep-th/0503138.

\bibitem{Aspinwall} P. Aspinwall and R. Kallosh, `` Fixing all moduli for M-theory
on $K3 \times K3$,'' hep-th/0506014. 

\bibitem{Kallosh2} E. Bergshoeff, R. Kallosh, A. Kashani-Poor, D. Sorokin 
and A. Tomasiello, ``An index for the Dirac operator on D3 brane with background 
fluxes,'' hep-th/0507069.

\bibitem{Witten2} E. Witten, ``Non-perturbative superpotentials in string theory,''
Nucl. Phys. {\bf B474} (1996) 343, hep-th/9604030.

\bibitem{Witten3} E. Witten, ``Small instantons in string theory,'' 
Nucl. Phys. {\bf B460} (1996) 541, hep-th/9511030.

\bibitem{Cederwall} M. Cederwall, ``Boundaries of eleven-dimensional membranes,''
Mod. Phys. Lett. {\bf A12} (1997) 2641, hep-th/9704161.

\bibitem{Sen} A. Sen, ``F theory and orientifolds,'' Nucl. Phys. {\bf B475} (1996) 562,
hep-th/9605150. 

\bibitem{Trivedi2} P. Tripathy and S. Trivedi, ``Compactification with flux on K3 
and tori,'' {\bf JHEP 0303} (2003) 028, hep-th/0301139.

\bibitem{Mattis} N. Dorey, T. Hollowood, V. Khoze and M. Mattis, ``Supersymmetry and 
multi-instanton measure 2. From N=4 to N=0,'' Nucl. Phys. {\bf B519} (1998) 470,
hep-th/9709072.

\bibitem{Mattis2} N. Dorey, T. Hollowood and V. Khoze , ``The calculus of many 
instantons,'' Phys. Rept. {\bf 371} (2002) 231, hep-th/0206063. 

\bibitem{Cherkis} S. Cherkis and J. Schwarz, ``Wrapping the M theory five-brane on K3,''
Phys. Lett. {\bf B403} (1997) 225, hep-th/9703062.

\bibitem{Hollowood} N. Dorey, T. Hollowood and S. Kumar, ``An exact 
elliptic superpotential for $N=1^*$ deformations of finite N=2 gauge theories,''
Nucl. Phys. {\bf B624} (2002) 95, hep-th/0108221.

\bibitem{Ovrut} E. Buchbinder, R. Donagi and B. Ovrut, ``Superpotentials for 
vectoe bundle moduli,'' Nucl. Phys. {\bf B653} (2003) 400, hep-th/0205190;
E. Buchbinder, R. Donagi and B. Ovrut, ``Vector bundle moduli superpotentials 
in heterotic superstrings and M theory,'' {\bf JHEP 0207} (2002) 066,
hep-th/0206203. 




\end{thebibliography}
\end{document}